\documentclass[sigconf,nonacm]{acmart}
\usepackage{tabularx}
\usepackage{subcaption}
\usepackage{enumitem}
\AtBeginDocument{%
  }

\renewcommand\footnotetextcopyrightpermission[1]{} % removes footnote with conference info
\setcopyright{none}
\setcopyright{acmlicensed}
\copyrightyear{2026}
\acmYear{2026}
\acmDOI{XXXXXXX.XXXXXXX}
\acmConference[KDD AI for Sciences Track]{}{August 9-13, 2026}{Jeju, Korea}
\begin{document}

\title{AgentCAT: An LLM Agent for Extracting and Analyzing Catalytic Reaction Data from Chemical Engineering Literature}

\author{Wei Yang$^1$, Zihao Liu$^1$, Tao Tan$^1$, Xiao Hu$^1$, Hong Xie$^{1_\ast}$, Lulu Li$^2$} 
\authornote{Hong Xie and Lulu Li are corresponding authors}
\author{Xin Li$^3$, Jianyu Han$^3$, Defu Lian$^1$, Mao Ye$^2$}
\affiliation{%
  \institution{$^1$University of Science and Technology of China, 
  State Key Laboratory of Cognitive Intelligence 
  \\
  $^2$National Engineering Research Center of Lower-Carbon Catalysis Technology,
   Dalian Institute of Chemical Physics, Chinese Academy of Sciences 
\\  $^3$IFLYTEK CO.LTD
  }
 \city{$^{1,3}$Hefei}
  \state{Anhui}
   \country{China}
     \city{$^{2}$Dalian}
  \state{Liaoning} 
  \country{China}
}
\renewcommand{\shortauthors}{Yang et al.}

\begin{abstract}
This paper presents a large language model (LLM) agent named AgentCAT, 
which extracts and analyzes catalytic reaction data from chemical engineering papers, 
%and supports natural language based interactive analysis of the extracted data.  
AgentCAT serves as an alternative to overcome the 
long-standing data bottleneck in chemical engineering field, 
and its natural language based interactive data analysis functionality is friendly to 
the community.  
AgentCAT also presents a formal abstraction and challenge analysis 
of the catalytic reaction data extraction task 
in an artificial intelligence-friendly manner.  
This abstraction would help the artificial intelligence community  
understand this problem and in turn would attract more attention to address it. 
Technically, the complex catalytic process leads to complicated dependency structure 
in catalytic reaction data  
with respect to elementary reaction steps, molecular behaviors, measurement evidence, 
etc.   
This dependency structure makes it challenging to 
guarantee the correctness and 
completeness of data extraction, as well as representing 
them for analysis.  
AgentCAT addresses this challenge and it makes four folds of technical contributions: 
(1) a schema-governed extraction pipeline with progressive schema evolution, enabling robust data extraction from chemical engineering papers;
(2) a dependency-aware reaction-network knowledge graph that links catalysts/active sites, synthesis-derived descriptors, mechanistic claims with evidence, and macroscopic outcomes, preserving process coupling and traceability;
(3) a general querying module that supports natural-language exploration and visualization over the constructed graph for cross-paper analysis;
(4) an evaluation on $\sim$800 peer-reviewed chemical engineering publications demonstrating the effectiveness of AgentCAT.
\end{abstract}

\maketitle

\section{Introduction}

Chemical engineering prioritizes engineering practicability and industrial scalability, 
which is quite different from pure chemical research that mainly  
focuses on reaction mechanisms 
\cite{Perry1950}.  
The core objective is 
translating lab-scale experimental feasibility into 
large-scale industrial production with commercial value, stable operation, 
and economic efficiency.  
Experiments at different scales, 
such as lab-scale screening, pilot scale validation, etc., 
form the factual basis to achieve that objective \cite{Zlokarnik2006}. 
The catalytic reaction data generated in these experiments 
reveals insights for understanding and optimizing the catalytic process.  

Reflecting on the history of chemical engineering, 
scarce catalytic reaction data is a long-standing bottleneck, 
which hinders the progress of this area.  
Several public datasets are constructed for the border chemistry area in recent years, such as ORD~\cite{Kearnes2021}, Orderly~\cite{Wigh2024}, and DigCat~\cite{Zhang2024}. These datasets typically represent reactions as isolated tuples and provide only limited insights for chemical engineering, where the catalytic process and reaction environments (especially scaled ones) distinguish reaction data from other subfields of chemistry. Experimental facts in chemical engineering are only interpretable when linked along the catalytic process, not as self-contained snippets.  
This leaves catalytic reactions reported in chemical engineering literature as the major data source. General-purpose LLMs such as Gemini and Grok excel in semantic understanding and multimodal reasoning, and serve as a promising alternative to extract such data from literature.

\begin{figure}[tb]
\centering
\includegraphics[scale=0.26]{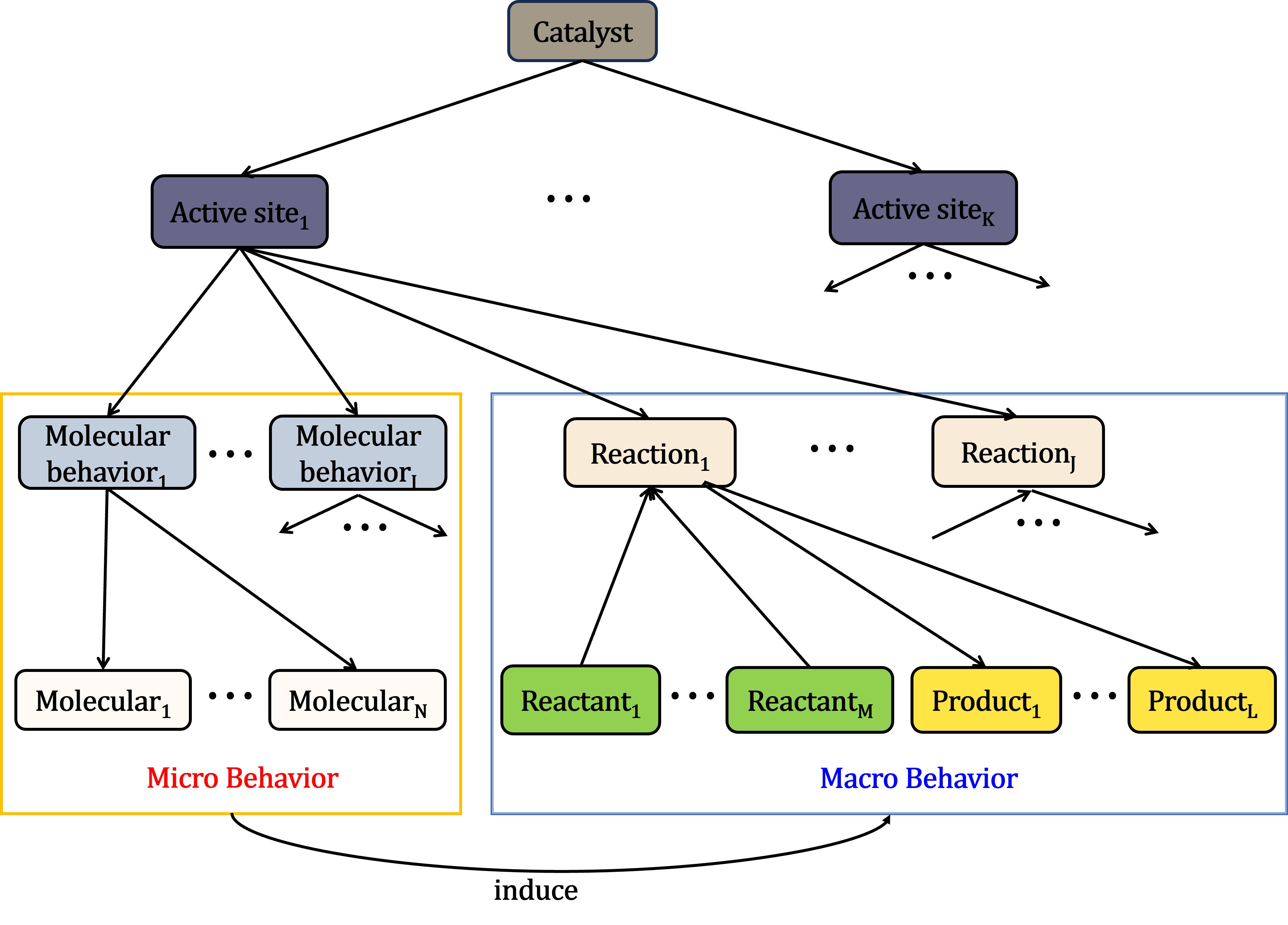}
\vspace{-0.12in}
\caption{
A graph abstraction of 
the backbone of the catalytic reaction data 
in chemical engineering.
}
\vspace{-0.22in}
\label{fig:ReactionHighLevel} 
\end{figure}

Before diving into technical details, let us get a glimpse of  
the catalytic reaction data in chemical engineering first.   
Figure \ref{fig:ReactionHighLevel} depicts a graph abstraction of 
the backbone of the catalytic reaction data 
(please refer to Section \ref{sec:preliminary} for more details).   
In Figure \ref{fig:ReactionHighLevel}, the top layer is a catalyst.  
Typical catalysts like molecular sieves, 
exhibit inherent complexity originating from their hierarchical structural features, 
tunable chemical properties, and dynamic physicochemical behaviors 
in practical applications.  
One typical catalyst is HZSM-5 \cite{Yang2019,Zhang2019}.  
Each catalyst has several active sites, where 
elementary reaction steps occur.  
For example, the active sites of HZSM-5 catalyst include 
Brønsted acid sites,
Lewis acid sites, 
defective hydroxyl sites, etc. 
The catalytic process can be formally described in micro level and macro level.  
The micro level characterizes molecular behaviors.  
For example, on Brønsted acid sites of HZSM-5, the 
main molecular behaviors include selective adsorption of reactant
molecules, subsequent bond cleavage/formation reactions of intermediates for product generation, desorption of products to regenerate the acid sites, etc.  
This micro level behavior induces the elementary reaction steps in the macro level.  
The catalytic process is usually highly unobservable.  
Techniques like probe molecular, 
pyridine-adsorbed IR spectra, etc., are employed to 
generate evidence of the catalytic process.  
These lines of evidence together with the tightly coupled elementary reaction steps  
form the catalytic reaction data.

While general-purpose LLMs such as Gemini and Grok excel 
in semantic understanding and multimodal reasoning, 
their direct application to extracting catalyst experimental data in chemical engineering often falls short in ensuring causal consistency, 
fine-grained alignment, and verifiable grounding. 
Scientific data extraction systems like 
SLM-MATRIX \cite{li_2025}, Eunomia \cite{ansari_2023} and nanoMINER \cite{odobesku_2025}
are domain-specific and they can not handle the aforementioned 
complex dependency structure in the catalytic reaction data.  
Our AgentCAT is motivated by a central observation in catalytic chemical engineering: catalyst experiment facts are only interpretable when they are complete, evidence-grounded, and linked along the catalytic process, rather than extracted as isolated reaction tuples. Accordingly, AgentCAT adopts a schema-governed and closed-loop design. We first make the extraction targets explicit via progressive schema evolution, then ground each field in traceable evidence through staged extraction and review-driven self-correction, and finally organize the extracted records into a dependency-aware reaction-network knowledge graph. Building on this graph, we support natural-language querying and visualization for interactive cross-paper exploration. The contributions:
\begin{itemize}[noitemsep,topsep=0pt,leftmargin=*] 
    \item We formulate chemical-engineering catalyst experiment extraction as reconstructing an SSP-oriented, process-coupled ``partial picture'' (mechanistic links, evidence, and experimental control logic), and identify three recurring failure modes of general LLMs in this setting: broken long-range causal linking, fine-grained parameter misalignment, and violations of domain logic without schema-level constraints.
    \item We propose a multi-agent, plan-then-execute pipeline featuring progressive schema evolution, a two-phase evidence-grounded extraction procedure, and review verdicts that trigger targeted re-extraction to improve completeness and reliability.
    \item  We design a reaction-network knowledge graph in Neo4j that links catalysts/active sites, synthesis-derived descriptors, mechanistic claims with evidence, and macroscopic outcomes with traceability to source PDFs, and we build a general querying agent with visualization to enable natural-language exploration across papers. We evaluate AgentCAT on $\sim$800 peer-reviewed chemical engineering publications and demonstrate robust end-to-end performance.
\end{itemize}

\vspace{-0.12in}
\section{Related Work}  

\noindent{
\bf Chemistry agent.} 
Recently, several LLM based chemistry agents are developed.  ChemCrow \cite{bran_2024} integrates an LLM with expert-designed chemistry tools to support multi-step problem solving, while CACTUS \cite{mcnaughton_2024} similarly connects LLMs to chem informatics tool chains for molecular property prediction, similarity search, and drug-likeness assessment. 
ChemAgent \cite{wu_2025} further scales up tool-augmented agent frameworks by aggregating diverse external chemistry tools, strengthening the end-to-end capability along the “tool selection–parameter filling–execution” pipeline.  
Several works investigate agentic systems for automated scientific information extraction from chemistry and materials literature. SLM-MATRIX \cite{li_2025} proposes a multi-agent reasoning and verification framework to improve the reliability of materials data extraction from papers. Eunomia \cite{ansari_2023} formulates a “chemist AI agent” that autonomously constructs materials datasets from scientific literature via planned extraction workflows. nanoMINER \cite{odobesku_2025} introduces a multi-agent multi modal extraction system for nano materials, highlighting the importance of handling heterogeneous evidence such as text, tables, and figures. Complementary to these agentic extraction systems, OpenChemIE \cite{fan_2024} provides a toolkit for document-level chemistry literature information extraction by integrating evidence across text, tables, and figures.  
However, these systems are domain-specific and can not 
handle the complex dependency structure in the catalytic reaction data.

\noindent{\bf Chemistry knowledge graph.}   
Chemistry knowledge graph (CKG) is a structured representation of chemistry entities, chemistry reactions, and their associated conditions \cite{wang2024}. 
%It plays an important role in both research and industrial applications. 
%Existing studies mainly focus on three types of methods. 
%There are three typical types of methods.  
%The first one is manual extraction method. 
Domain experts manually analyze literature and extract reactants, products, and reaction conditions, such as BioGRID \cite{oughtred2021}. 
These approaches can achieve high-precision knowledge extraction and ensure the accuracy of entities and relationships, but they are time-consuming, labor-intensive, and difficult to scale to large volumes of literature.  
Rule-based or shallow model methods analyze literature using rule-based systems or shallow machine learning models, enabling rapid processing of large-scale literature and partially automating knowledge graph construction, thus significantly improving efficiency \cite{lin2020, nguyen2018, xia2023, zhou2022}. However, chemistry literature often contains a mixture of natural language, chemistry symbols, tables, and implicit descriptions, which limits the performance of these methods in complex semantic understanding and cross-sentence dependency handling, thereby affecting the robustness and accuracy of extraction. 
LLMs-based methods leverage the powerful contextual semantic understanding of LLMs, these methods can automatically extract entities and relationships, supporting knowledge graph construction in materials and other domains, and achieving preliminary large-scale automated extraction \cite{ bian2025,omar2025, yoshitake2025}. Nevertheless, most of these methods are restricted to specific domains or task settings, and a general, fully automated construction of CKG remains an open challenge. 
To address the limitations of these existing methods, this paper builds 
AgentCAT. AgentCAT combines the semantic understanding capabilities of the Gemini LLM with automated extraction techniques, enabling efficient and accurate processing of complex literature.  

\noindent{\bf LLM-based knowledge graph construction.}  
%In recent years, 
Large language models (LLMs) have transformed knowledge graph (KG) construction \cite{mo2025kggen}. 
Early approaches like Microsoft’s GraphRAG extracts triples at the sentence or document level but often produced sparse, fragmented graphs with poor connectivity and limited downstream performance \cite{edge2024graphrag}. 
A major advance is introduced in KGGen \cite{mo2025kggen}, which combines LLM-driven triple extraction with advanced post-processing techniques. 
It effectively resolves duplicate entities and consolidates relations, resulting in denser and more coherent knowledge structures compared to prior extractors.
Recent progress includes agentic frameworks that enable autonomous planning, self-correction, and ontology-aligned construction \cite{chen2025knobuilder,mohammadi2025llmagentsurvey}, as well as production tools like Neo4j’s LLM KG Builder supporting multi-model integration and parallel processing \cite{hunger2025llmkgbuilder}. However, these methods still lack strong domain-specific prior knowledge (e.g., chemical validity checks), struggle to maintain global consistency across large-scale corpora, and have limited validation on downstream tasks such as reaction prediction \cite{wang2024,bian2025,li2025llmkgconstruction,yoshitake2025}.

\section{Preliminary\&Challenges}
\label{sec:preliminary}

%We first present a graph abstraction of the reaction data in 
%chemical engineering to illustrate the inherent structure complexity. 
%Then we use concrete examples to show the challenges of using 
%LLMs to extract such reaction data from chemical engineering papers.  

\subsection{Graph Abstraction of Catalytic Reaction Data in Chemical Engineering}
\label{sec:wet_experiments}

\begin{figure}[tb]
\centering
\includegraphics[scale=0.26 ]{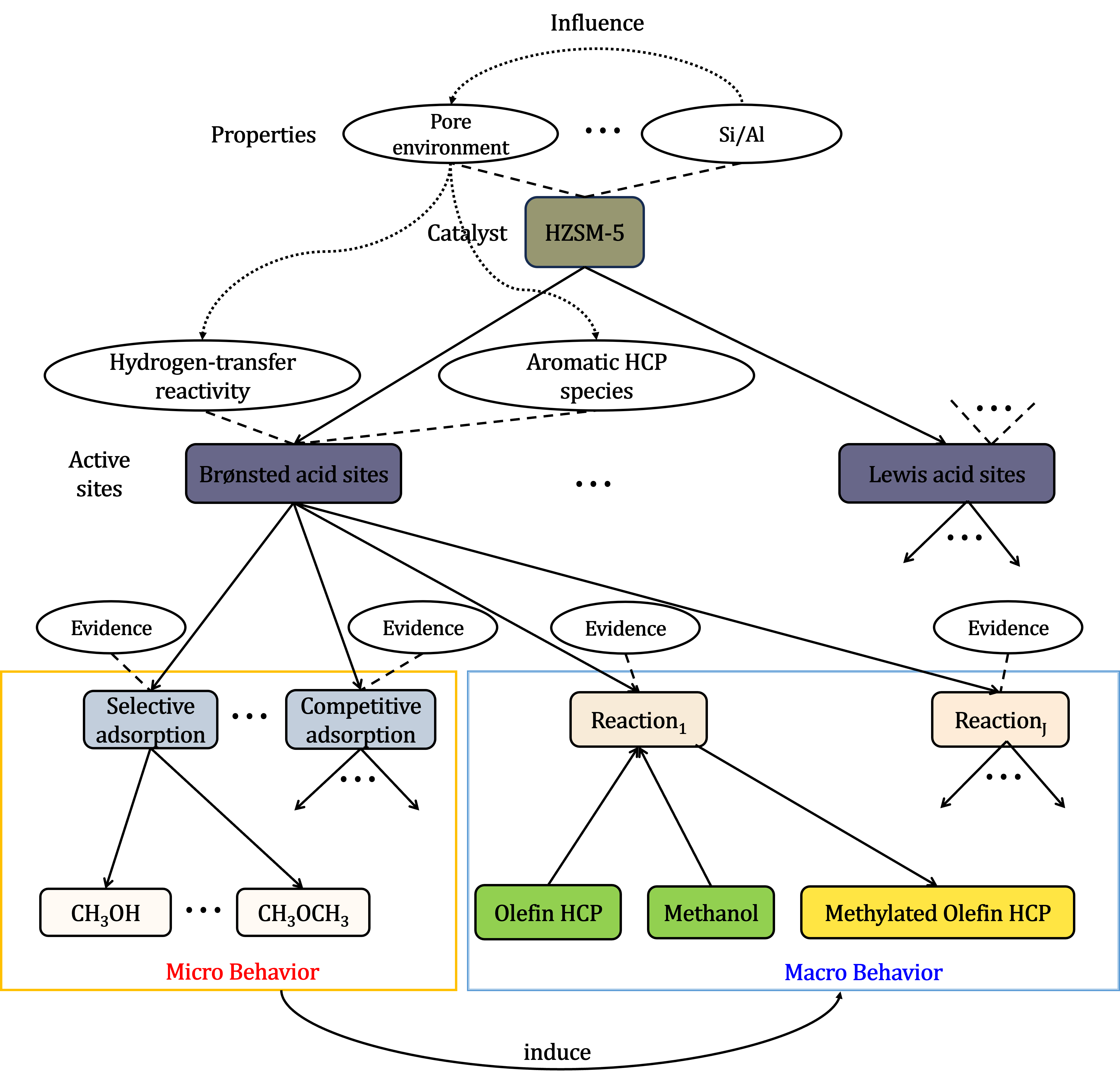}
\vspace{-0.16in}
\caption{
A simplified SSP-oriented schematic of how microscopic mechanistic events governed by catalyst active sites and the pore environment propagate to macroscopic reaction outcomes in H-ZSM-5 catalyzed MTO.
}  
\vspace{-0.2in}
\label{fig:ReactionExample} 
\end{figure}

In chemical engineering, reactions are rarely isolated events, 
but instead they are tightly coupled by the catalytic process.  
Experimental facts in this domain are only interpretable when 
they are linked along this catalytic process, 
rather than treated as self-contained snippets.  
The catalytic process is usually highly unobservable.  
Techniques like probe molecular, 
pyridine-adsorbed IR spectra, etc., are employed to 
generate evidence of the catalytic process.  
These lines of evidence together with the tightly coupled reactions 
form a ``partial picture'' of the catalytic process.  
This partial picture, though not full, 
serves as essential factual basis and 
reveals key insights to optimize chemical engineering experiment design.  
The objective is to extract this type of ``partial picture'' from 
chemical engineering papers.  
Figure \ref{fig:ReactionExample} depicts a graph abstraction of 
this ``partial picture''.  
To strike a balance between concreteness and readability, 
Figure \ref{fig:ReactionExample} uses the 
Methanol-to-Olefins (MTO) reaction \cite{Yang2019,Zhang2019}
as an example.  
Note that MTO is a typical chemical engineering reaction.   
Essentially, it is catalytic process that converts methanol into light olefins  
using acid zeotype catalysts in reactors. 
Figure \ref{fig:ReactionExample} represents 
the \textit{Synthesis--Structure--Performance} (SSP) paradigm: 
synthesis protocols determine physicochemical structures, which in turn constrain mechanistic pathways and ultimately manifest as observable performance metrics. 
Consequently, experimental facts in this domain are only interpretable when they are linked along this causal lineage, rather than treated as self-contained snippets.  
Figure \ref{fig:ReactionExample} depicts a graph that abstracts key elements of this partial picture.  
One can observe that 
synthesis-controlled descriptors (e.g., Si/Al $\rightarrow$ acidity) 
determine active site-level events (e.g., hydrogen-transfer propensity and the formation/retention of aromatic HCP species), 
which then propagate to macroscopic observables such as 
reactants (Olefin HCP and Methanol) and products (Methylated Olefin HCP). 
We next introduce necessary details on the ``complexity'' of the catalytic process and data.  

\noindent
{\bf Dependency on material synthesis.}   
Many factors that influence the catalytic process.   
Take the MTO study reported in \cite{Liang2023Insight} as an example.   
The authors prepare a series of H-ZSM-5 zeolites denoted as TZ-$m$, 
where $m$ corresponds to the Si/Al ratio, spanning approximately 22 to 195. 
Within this controlled catalyst family, changing Si/Al systematically alters acidity-related descriptors (e.g., acid amount/density and Al-site configurations such as paired vs.\ single Al), which then impacts hydrocarbon-pool (HCP) evolution and the balance between competing catalytic cycles.  
Therefore, extracting a performance metric such as ``ethene selectivity'' is scientifically incomplete unless it is explicitly anchored to the catalyst identity (TZ-$m$) and its synthesis-derived structural descriptors (e.g., Si/Al and acidity descriptors) defined elsewhere in the paper \cite{Liang2023Insight}.  

\noindent
{\bf Mechanism-driven complexity beyond macroscopic I/O.} 
A  catalytic reaction encapsulates not only macroscopic reactant--product outcomes but also microscopic mechanistic events governed by catalyst active sites and the pore environment. In the example paper \cite{Liang2023Insight}, 
acidity-related factors modulate hydrogen-transfer reactivity and the buildup of retained/aromatic HCP species (including polymethylbenzenes), and these microscopic trends are connected to differences in macroscopic product distributions \cite{Liang2023Insight}.  
Notably, it is emphasized in \cite{Liang2023Insight} that, under pulse-reactor comparisons, the catalyst bed mass is adjusted to keep the total Br{\o}nsted acid site (BAS) amount comparable across samples with different BAS densities; this control logic is essential for interpreting downstream comparisons of conversion and selectivity.  

\noindent
{\bf Evidence of the catalytic process.} 
%  sketches this SSP chain for the H-ZSM-5 case: synthesis-controlled descriptors (e.g., Si/Al $\rightarrow$ acidity) determine site-level events (e.g., hydrogen-transfer propensity and the formation/retention of aromatic HCP species), which then propagate to macroscopic observables such as olefin distributions and stability-related trends \cite{Liang2023Insight}.  
The mechanistic links in Figure~\ref{fig:ReactionExample} are not asserted as standalone statements; they are supported by heterogeneous evidence. Specifically, the study employs $^{13}$C/$^{12}$C methanol switching experiments and $^{13}$C-methanol/$^{12}$C-xylene co-feeding experiments to probe the origin of olefins and the role of aromatic HCP species (e.g., whether olefins arise predominantly from aromatics cracking versus alternative pathways)  \cite{Liang2023Insight}.  

\noindent
{\bf Heterogeneity in experimental design.} 
Experimental outcomes in chemical engineering are highly sensitive to design choices, which creates a high-dimensional parameter space and makes standardization non-trivial. Even within a single paper, the \emph{same} nominal reaction can be interrogated under distinct operational regimes with fundamentally different semantics. 
The example study \cite{Liang2023Insight} employs (i) a pulse micro-reactor to probe transient behaviors during the induction period and to analyze retained HCP species under controlled dosing, and (ii) a continuous-flow fixed-bed reactor to evaluate steady-state performance and catalyst stability under extended operation (reported with explicit space-velocity settings). These setups answer different scientific questions (transient mechanism probing vs.\ lifetime evaluation). Therefore, extracted measurements such as ``conversion'' or ``selectivity'' must be contextualized by reactor mode, temporal regime (e.g., pulse index vs.\ time-on-stream), and the associated protocol; otherwise, merging values across modes leads to invalid comparisons.  
A particularly challenging but common design pattern is \emph{controlled comparison}, where authors intentionally normalize one factor to isolate another. In the pulse tests of the example paper, catalyst bed mass is adjusted to keep the total Br{\o}nsted acid site (BAS) amount comparable across catalysts with different acid site densities. Such normalization is critical for correctly interpreting downstream differences in product distributions and HCP evolution. Accordingly, an extraction system must represent not only the reported outcomes but also the experimental control logic (what is held constant and how), since this logic determines what cross-sample comparisons are valid.

\noindent
{\bf Summary.}
An effective data extraction system must therefore recover mechanistic claims \emph{together with} their supporting evidence and experimental context, rather than only extracting surface-level entities (reactants, products, and yields).  

\begin{table*}[ht] % [t] 表示优先置于页面顶部，table* 用于跨双栏
    \centering
    \caption{Representative extraction discrepancies observed in Grok-3 outputs on processing \cite{Liang2023Insight}. Each row contrasts the erroneous extraction with the original context and domain logic, illustrating the three failure modes identified in Section~\ref{sec:challenges}.}
\label{tab:extraction_discrepancies}
\begin{tabularx}{\textwidth}{
    p{0.12\textwidth}                         % 第一列保持 15%
    >{\raggedright\arraybackslash\hsize=0.60\hsize}X  % 第二列：系数 0.8
    >{\raggedright\arraybackslash\hsize=1.40\hsize}X  % 第三列：系数 1.2
}
    \toprule
    \textbf{Category} & \textbf{Grok3 Extraction Error} & \textbf{Original Context \& Domain Logic} \\
    \midrule

    % Category 1
    \multicolumn{2}{l}{\textbf{1. Lost-in-the-middle in Causal Linking}} \\
    \midrule
    \textit{Experimental Control Causality} & 
    Records \texttt{"loading\_adjustment"} to unify BAS amount but fails to link it to the outcome. & 
    \textbf{Original Logic:} The adjustment of catalyst weight \textit{causes} the similar methanol conversion ($\sim$24\%) across samples, proving the reaction is BAS-catalyzed. The extraction isolates the method from its intended validation result. \\
    \midrule

    % Category 2
    \multicolumn{3}{l}{\textbf{2. Hallucination in Fine-Grained Parameter Alignment}} \\
    \midrule
    \textit{Characterization Data Misalignment} & 
    Attributes NH$_3$-TPD acid amounts (e.g., Weak: 247 $\mu$mol/g) to \texttt{"methanol"} adsorption site strength. & 
    \textbf{Hallucination:} The values (247, 203 $\mu$mol/g) are specific to NH$_3$ desorption peaks (Table 2). Assigning them to methanol adsorption creates a phantom property not measured in the paper. \\
    \addlinespace
    \textit{Pulse Parameter Ambiguity} & 
    Generalizes pulse numbers as \texttt{"11-15 pulses"} in \texttt{reaction\_system}. & 
    \textbf{Precision Error:} The paper distinguishes strict protocols: 11 pulses for retained species analysis (in paper's Fig. 5) vs. $14+4$ pulses for isotopic switching. Merging them prevents accurate replication of experimental stages. \\
    \midrule

    % Category 3
    \multicolumn{3}{l}{\textbf{3. Lack of Domain Logic Constraints}} \\
    \midrule
    \textit{Property Attribution Violation} & 
    Defines \texttt{site\_density} of \texttt{"aromatic HCP species"} as 131 $\mu$mol/g. & 
    \textbf{Domain Logic Violation:} 131 $\mu$mol/g is the density of Brønsted Acid Sites on the zeolite framework (TZ-20). It is physically invalid to assign a catalyst's structural property as the density of a transient reaction intermediate. \\
    \addlinespace
    \textit{Missing Steric Constraints} & 
    Describes aromatic elimination steps without noting spatial confinement. & 
    \textbf{Mechanism Loss:} The paper argues that bulky aromatics (TetraMB-HexaMB) are spatially trapped in intersections, limiting their access to acid sites. The extraction misses this geometric constraint, which is central to explaining the "independence from acid density." \\
    \bottomrule
    \end{tabularx}
\end{table*}

\subsection{Challenges for General LLMs}
\label{sec:challenges}
While general-purpose LLMs such as Gemini and Grok excel in semantic understanding and multimodal reasoning, their direct application to extracting catalyst experimental data in chemical engineering often falls short in ensuring causal consistency, fine-grained alignment, and verifiable grounding. Through a representative case study on H-ZSM-5 catalysis~\cite{Liang2023Insight}, we identify three recurring failure modes, with representative examples listed in Table~\ref{tab:extraction_discrepancies}.  
We elaborate on these errors as follows.  

\noindent
{\bf Lost-in-the-middle in causal linking.}
Chemical engineering papers distribute dependent information across distant sections and modalities: synthesis and control logic appear in experimental protocols, characterization data in separate results sections, and mechanistic interpretations elsewhere. Correct extraction requires maintaining long-range causal links among these distributed elements. Standard LLMs frequently break these dependencies or conflate attributes across nearby sample descriptions, producing internally inconsistent records.

\noindent
{\bf Hallucination in fine-grained parameter alignment.}
Many reported quantities are not static catalyst properties but protocol- and time-indexed measurements. Isotopic switching experiments, pulse sequences, and co-feeding designs introduce temporal structure (e.g., stage ordering and mode-specific readouts). When an LLM extracts a value without preserving its experimental context, the number loses physical meaning and becomes unusable for downstream analysis.

\noindent
{\bf Lack of domain logic constraints and schema-level verification.}
catalyst experiment data must satisfy domain constraints induced by chemistry and experimental design. For instance, in zeolite series such as TZ-$m$, changes in Si/Al correlate with systematic shifts in acidity-related properties; controlled comparisons further impose constraints on valid attribute pairings. A free-form LLM has no built-in mechanism to enforce these constraints or to detect violations such as assigning a catalyst's structural property to a transient reaction intermediate.

\begin{figure*}[ht]
\centering
\includegraphics[width=0.975\linewidth]{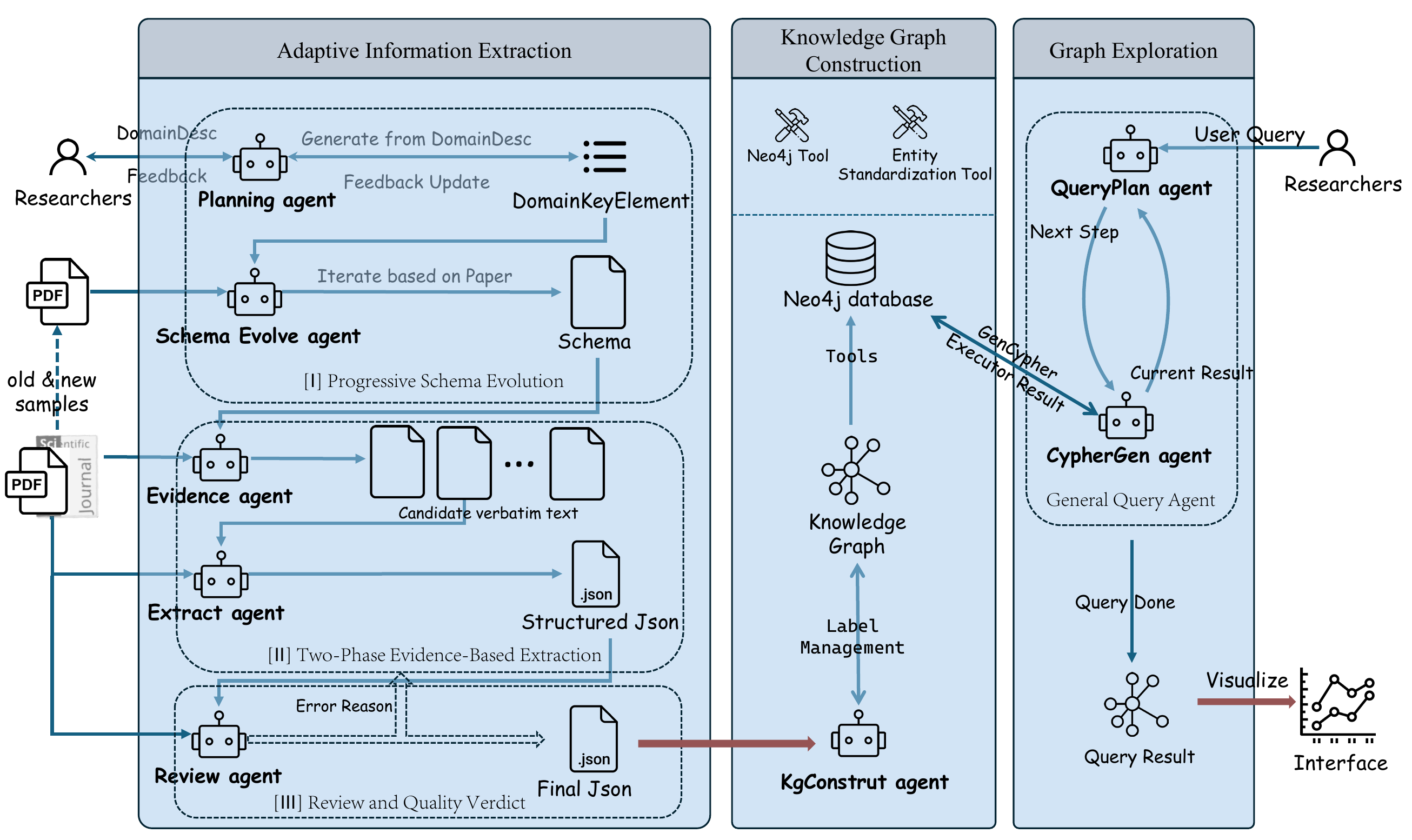}
\caption{
Architectural overview of AgentCAT, showing the multi-phase process of converting raw PDF literature into structured data and constructing a Knowledge Graph through adaptive extraction, schema evolution, and dynamic graph exploration.
}
\label{fig:archi}
\end{figure*}

\section{The Design of AgentCAT}
\label{sec:design}
\subsection{Architectural Overview}
\label{sec:design_philosophy}
To address the challenges outlined in Section~\ref{sec:challenges}, AgentCAT is designed as a multi-agent orchestration framework governed by two fundamental principles: (i) plan-then-execute and (ii) closed-loop self-correction. Specifically, the system decouples the extraction workflow into distinct phases—schema specification, evidence collection, and verification—mirroring the rigorous methodology of human researchers who iteratively refine an initial plan through execution and validation cycles. Meanwhile, feedback loops are embedded at every layer: from JSON parsing at the infrastructure level, to semantic review at the validation layer, to query repair during graph exploration—ensuring errors are detected and corrected in situ rather than propagating downstream.

As illustrated in Figure~\ref{fig:archi}, this design is instantiated through three progressive stages: Adaptive Extraction, Knowledge Graph Construction, and Interactive Graph Exploration. Together, these stages directly target the three failure modes identified in Section~\ref{sec:challenges}: progressive schema evolution and dependency-aware graph construction preserve long-range causal links across dispersed sections; candidate-then-resolve evidence grounding improves fine-grained parameter alignment; and review-driven closed-loop re-extraction enforces domain-logic consistency.

\subsection{Adaptive Information Extraction}
\label{sec:extraction}

At the core of AgentCAT lies its ability to extract high-fidelity, structured data from complex and heterogeneous scientific texts. We implement a multi-stage extraction pipeline, orchestrated by LangGraph, to address the challenges of long-range dependencies and hallucination in fine-grained parameter alignment.

\noindent
{\bf Progressive schema evolution.}
Prior approaches to literature structuring rely on domain researchers to manually design static schemas. This paradigm heavily depends on both the researcher's ontology modeling expertise and their proficiency in schema specification (e.g., JSON schema design), and often fails to capture the nuances of domain-specific knowledge, particularly in highly specialized areas such as catalytic chemical engineering. To overcome these limitations, we introduce a Human-in-the-Loop Planning mechanism. Initially, the Planning Agent collaborates with researchers to establish a preliminary \texttt{DomainKeyElements}, which serves as the schema seed. Subsequently, the Schema-Evolution Agent iteratively processes a stratified sample of reference papers, which spanning both historical and recent publications,  to refining the schema by incorporating new entity types or hierarchical properties while ensuring backward compatibility. This dynamic schema evolution allows AgentCAT to adapt to emerging research paradigms, while still retaining the ability to reprocess legacy documents.

\noindent
{\bf Two-phase evidence-based extraction.}
To enhance extraction granularity, we partition the schema into multiple segments and process each through a ``Candidate-then-Resolve'' strategy, mitigating the risk of hallucination and ensuring that extracted information is grounded in textual evidence:
\begin{itemize}[noitemsep,topsep=0pt,leftmargin=*] 
    \item \textbf{Phase 1 (Candidate):} The agent scans the document section by section to extract verbatim text fragments as potential evidence for specific schema fields, without any interpretation or transformation.
    \item \textbf{Phase 2 (Resolve):} These verbatim candidates are then processed alongside their original context to populate the final structured fields. This separation compels the model to ground its outputs in explicit textual evidence, thereby enhancing the quality and traceability of the extraction.
\end{itemize}

\noindent{\bf Review and quality verdict.}
Extracted data undergoes rigorous review by the review agent. The review process yields one of three verdicts: \texttt{PASS}, \texttt{MINOR\_FIX} (for formatting issues), or \texttt{MAJOR\_ERROR}. In the case of a \texttt{MAJOR\_ERROR}, a re-extraction loop is triggered, during which error-reason hints are injected back into the extraction prompt to dynamically guide the agent toward more accurate outputs in subsequent iterations. This mechanism ensures continuous improvement throughout the extraction process.

\subsection{Knowledge Graph Construction}
\label{sec:kg_construction}
Different chemical research domains require tailored knowledge graphs. In this work, we focus on the interaction between zeolite catalysts and reaction processes within the chemical engineering domain, linking them through active sites as the central bridge. Accordingly, AgentCAT's Representation Module transforms the structured JSON data produced by the Extraction Module into a Reaction Network Graph stored in a Neo4j graph database, which can be adapted to other graph structures as needed. This process explicitly models the relationships among catalyst properties, microscopic molecular adsorption, and macroscopic catalytic processes, addressing the challenge of ``Reaction Dependency'' by capturing the complex interdependencies across chemical reactions.

\noindent{\bf Dynamic labeling management.}
Scientific discovery frequently introduces novel concepts that cannot be captured by predefined ontologies, while identical concepts may appear under varying nomenclatures. Unlike traditional knowledge graphs with static schemas, our Label Manager supports dynamic schema expansion under a conservative policy, where new labels are introduced only when they are strictly necessary. This restraint preserves structural coherence of the graph. For instance, if the extraction process identifies a novel catalyst property or reaction condition, the system dynamically registers new node labels in the graph database. This enables the knowledge graph to evolve continuously while maintaining its structural integrity as the scientific literature progresses.

\noindent{\bf Entity normalization and linking.}
To ensure consistency and connectivity within the graph, we employ a \textbf{Molecular Normalizer} that standardizes chemical formulas (e.g., unifying ``propylene'', ``propene'', and ``$C_3H_6$'') and resolves naming discrepancies in catalyst nomenclature. Additionally, essential nodes in the graph are linked to their source PDF identifiers, providing traceability and facilitating verification of graph-based information against the original literature.

\subsection{General Querying and Graph Exploration}
\label{sec:reasoning}

The knowledge graph produced by AgentCAT is designed to assist researchers in querying and interacting with the reaction network, facilitating their understanding of catalytic mechanisms and providing a cross-literature perspective that may inspire novel research directions.

\noindent{\bf General querying agent for simplified exploration.}
Researchers' querying needs for the knowledge graph stored in the Neo4j database are often diverse and unpredictable. Fixed query templates fail to accommodate this variability, while expecting researchers to directly manipulate the database imposes an unreasonable burden. To address this, we design a General-Querying Agent that employs a natural language interface, enabling researchers to interact with the reaction network intuitively rather than being constrained by raw data or structured query languages. For instance, a researcher might query, ``Which catalysts are designed for producing product X?'' The agent progressively decomposes such complex queries into a detailed query plan—determining which sub-queries to execute first, synthesizing intermediate results, revising the plan as needed, and deciding subsequent query steps. Finally, the agent consolidates the complete results and returns them to the visualization interface, offering researchers an intuitive understanding of the network.

\noindent{\bf Graph visualization for knowledge discovery.}
The querying agent is integrated with a graph visualization interface that enables researchers to intuitively trace and comprehend the details of catalytic reactions—insights that are difficult to obtain from plain text alone. Crucially, this interface unifies the catalytic mechanisms from multiple publications into a single visual canvas, empowering researchers to explore and query across the boundaries of individual papers. This cross-document view highlights both the similarities and differences in catalytic patterns, reveals non-obvious research directions, and provides anchor points for connecting designs across studies, which ultimately enhances scientific reasoning and accelerates knowledge discovery.

\section{Performance Evaluation}
\subsection{Evaluation Settings}
\noindent{\bf Dataset.}
We evaluate AgentCAT on $\sim$800 peer-reviewed publications from the energy catalysis domain within chemical engineering. This dataset was curated and validated by domain experts, who confirmed that these papers collectively represent the core findings and methodological breadth of the field, thereby constituting a representative benchmark for agent evaluation.

\noindent{\bf Implementation details.}
We implement all agents using LangGraph. For \textit{adaptive information extraction}, we employ
\texttt{Gemini\allowbreak-2.5\allowbreak-Pro} via Google AI Studio. For \texttt{knowledge graph construction}
and \textit{general querying \& graph exploration}, we adopt \texttt{doubao\allowbreak-seed\allowbreak-1.8} via VolcanoArk to
better support Chinese-language content and multilingual research workflows. The constructed knowledge graph is persisted in
\texttt{Neo4j} database.

\subsection{Performance of the Schema Evolution}
\label{sec:sch_evo}
\begin{figure}[b]
    \centering
    \includegraphics[width=0.90\linewidth]{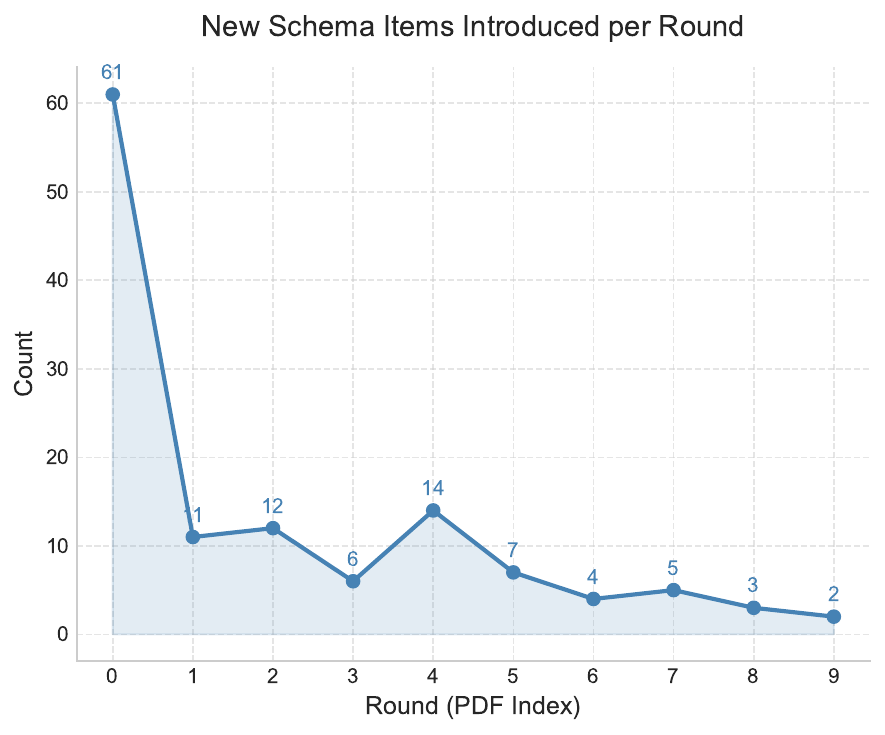}
    \caption{Number of newly introduced schema items per evolution round (each round processes one PDF).}
    \label{fig:schema_evo_new_items}
\end{figure}

We evaluate the convergence behavior of our schema evolution using 10 representative PDFs. Starting from an empty schema, we process documents sequentially and track the count of newly induced items (entities, attributes, or properties) per round. Figure~\ref{fig:schema_evo_new_items} reveals a bootstrapping-then-convergence pattern: the initial round establishes the core structure, while subsequent rounds contribute only marginal extensions to accommodate specific reporting variances. This rapid saturation confirms that our schema quickly stabilizes, requiring minimal updates for processing further literature.

\subsection{Performance of Data Extracting}
\label{sec:53}
\begin{figure*}[ht]   
  \centering
  \vskip 0.2in
  \begin{subfigure}[t]{0.35\textwidth}
    \centering
    \includegraphics[width=\linewidth]{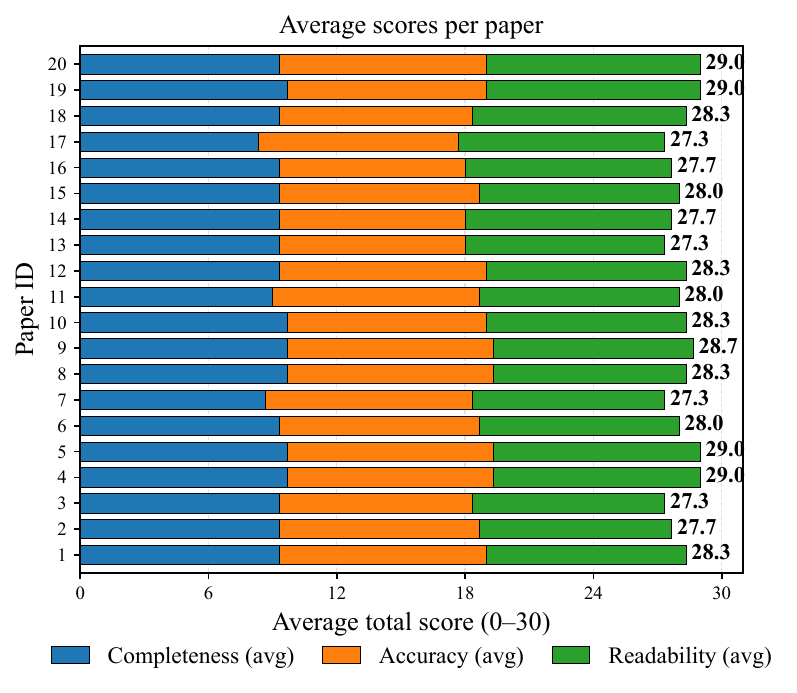}
    \caption{Expert blind scores: Completeness, Accuracy, Readability}
    \label{fig:paper_scores.png}
  \end{subfigure}
  \hfill
  \begin{subfigure}[t]{0.37\textwidth}
    \centering
    \includegraphics[width=\linewidth]{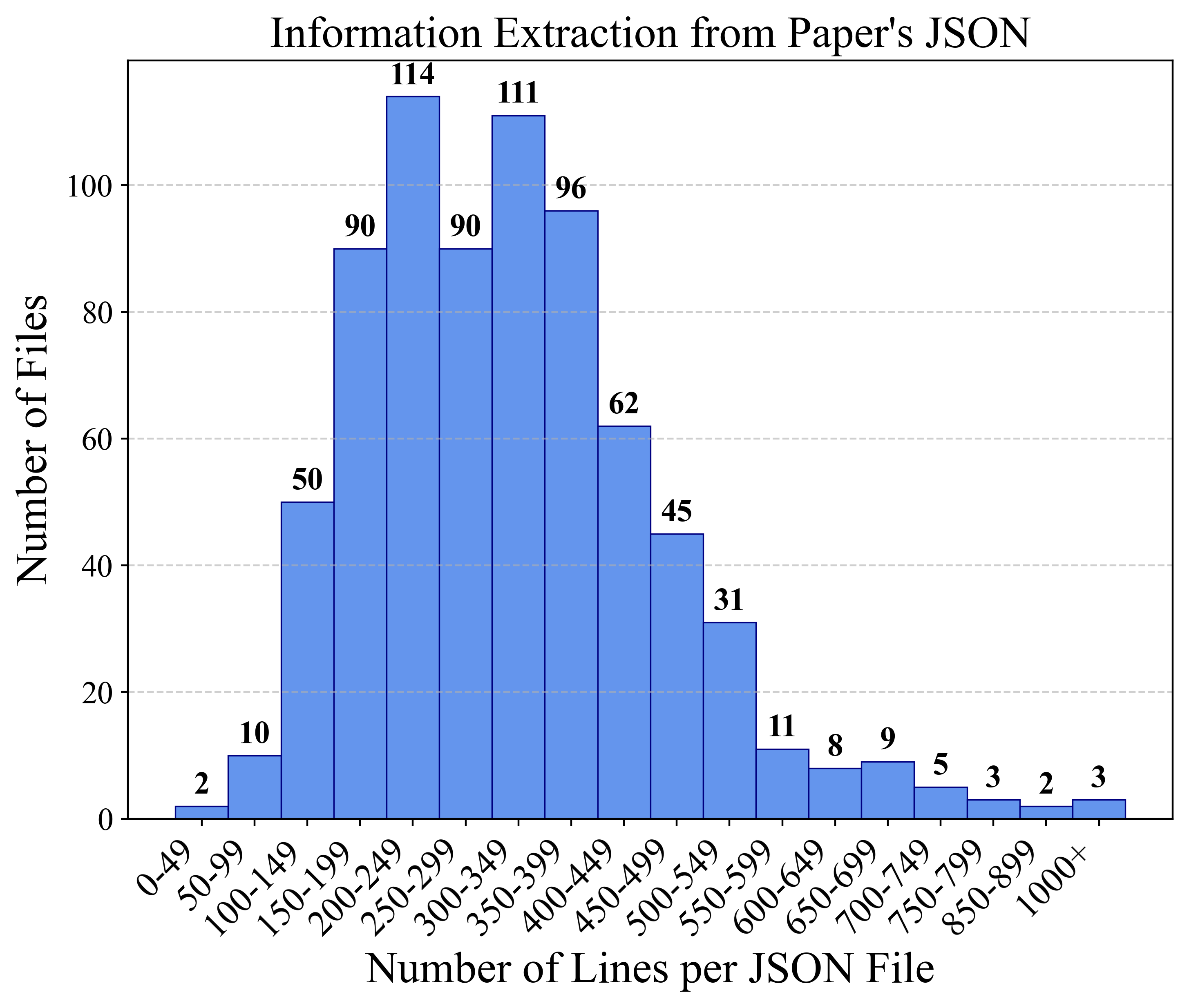}
    \caption{JSON length distribution (lines) over 733 papers}
    \label{fig:json_line_count}
  \end{subfigure}
  \hfill
  \begin{subfigure}[t]{0.26\textwidth}
    \centering
    \includegraphics[width=\linewidth]{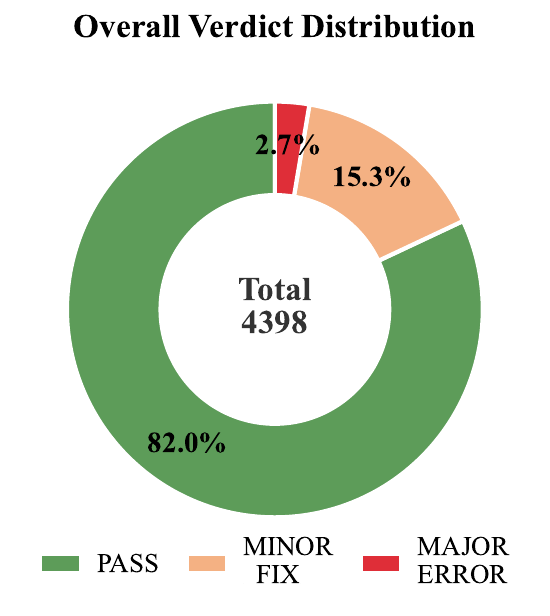}
    \caption{Review verdict distribution}
    \label{fig:pie}
  \end{subfigure}
  \caption{Extraction quality and reliability of AgentCAT on a large catalysis corpus. (a) Distribution of extracted JSON length as a coarse proxy for information density. (b) Expert blind ratings of extracted outputs across Completeness/Accuracy/Readability . (c) Overall review-verdict distribution over all extracted sections.}

  \label{fig:figure5.3}
\end{figure*}

\noindent
{\bf Chemical engineering expert evaluation.} 
Evaluating the completeness and accuracy of extraction is 
challenging for LLMs. Thus, we employ three chemical 
engineering experts to evaluate the extraction of 
twenty randomly selected papers.  The experts assessed each extracted record along three axes: \textit{Completeness}, \textit{Accuracy}, and \textit{Readability}, each scored from 1 to 10. Figure ~\ref{fig:paper_scores.png} visualizes the averaged scores across the 20 sampled papers. Across nearly all cases, \textit{Accuracy} and \textit{Readability} are consistently strong, while \textit{Completeness} shows modest variation but remains robust overall. These results suggest that AgentCAT produces structured records that are easy to inspect, largely consistent with the source, and comprehensive for downstream analysis. Qualitative feedback and detailed comments on nine of the papers are summarized in Appendix~\ref{apd:result} (Table~\ref{tab:Expert Evaluations}); these comments also serve as guidance for future refinements.

\noindent
{\bf Computing-level performance.}
Figure~\ref{fig:json_line_count} reports the distribution of JSON output lengths for 733 catalysis papers processed by AgentCAT. While a coarse and format-dependent metric, line count provides a rough proxy for the richness within the extracted dataset. 
We further assess reliability at scale by aggregating review verdicts over the adaptive extraction phase. Across 4,398 extracted sections from 733 papers, the review agent assigns one of three labels: \texttt{PASS}, \texttt{MINOR\_FIX}, or \texttt{MAJOR\_ERROR} (Figure~\ref{fig:pie}). The pipeline achieves an 82.0\% \texttt{PASS} rate, with 15.3\% \texttt{MINOR\_FIX} and only 2.7\% \texttt{MAJOR\_ERROR}. We also break down the \texttt{MAJOR\_ERROR} cases across the six common section types in Appendix~\ref{apd:result} (Figure~\ref{fig:majorerror}), which helps identify where hard failures concentrate and can guide future improvements. The low incidence of major failures supports the value of review-driven gating, while the non-trivial \texttt{MINOR\_FIX} fraction suggests that improving `single-run-pass' acceptance remains an important optimization target. Overall, the current extraction phase delivers structured data with practical fidelity that can be readily consumed by subsequent knowledge-graph construction and deeper catalysis studies.
\subsection{Performance of General Querying Agent}
To evaluate the robustness of our General Querying Agent built on the \texttt{doubao-seed-1.8} model, we designed a benchmark comprising 12 queries across three difficulty levels (\textit{Easy}, \textit{Medium}, and \textit{Hard}, four queries each). Each query was executed five times (60 trials in total) to account for the stochastic nature of LLM generation. We measured the error rate and average execution time per query.
As shown in Table~\ref{tab:query_performance}, the agent achieves an overall correctness rate of 86.67\%. Both \textit{Easy} and \textit{Medium} queries exhibit a 10\% error rate, while \textit{Hard} queries reach 20\%, indicating graceful degradation on complex reasoning tasks. Notably, execution time increases with difficulty: \textit{Medium} queries require approximately 55\% more time than \textit{Easy} queries (149.17s vs.\ 96.36s) despite identical error rates. However, the relatively modest time gap between these two levels suggests that the planning phase rather than Cypher execution dominates overall latency.Furthermore, even \textit{Easy} queries incur latencies substantially higher than conventional database queries. While slow inference speed of \texttt{doubao-seed-1.8} partly accounts for this overhead, these observations point to considerable room for improvement in both plan generation and query decomposition to better accommodate the interactive needs of researchers.

\begin{table}[hb]
\centering
\caption{Performance of the General Querying Agent across difficulty levels, categorized by the number of hops in query logic. All queries were executed successfully; error rate denotes cases where returned results were incomplete (e.g., missing nodes or relationships).}
\begin{tabular}{lcc}
\toprule
\textbf{Difficulty} & \textbf{Error Rate (\%)} & \textbf{Avg.\ Time (s)} \\
\midrule
Easy   & 10.00 & 96.36  \\
Medium & 10.00 & 149.17 \\
Hard   & 20.00 & 308.32 \\
\midrule
\textbf{Overall} & \textbf{13.33} & \textbf{184.62} \\
\bottomrule
\end{tabular}
\label{tab:query_performance}
\end{table}

\subsection{System Interface} 
Appendix~\ref{apd:sys_demostrat} presents example queries using the \textit{general query} feature and corresponding results, plus sample logs showing query decomposition.

\section{Conclusion}  
This paper presents the design and anlysis of AgentCAT.  
AgentCAT extracts and analyzes catalytic reaction data from chemical engineering papers, 
%and supports natural language based interactive analysis of the extracted data.  
AgentCAT serves as an alternative to overcome the 
long-standing data bottleneck in chemical engineering field.  
AgentCAT also presents a formal abstraction and challenge analysis 
of the catalytic reaction data extraction task 
in an artificial intelligence-friendly manner.   
AgentCAT makes four folds of technical contributions: 
(1) a schema-governed extraction pipeline with progressive schema evolution, enabling robust data extraction from chemical engineering papers;
(2) a dependency-aware reaction-network knowledge graph that links catalysts/active sites, synthesis-derived descriptors, mechanistic claims with evidence, and macroscopic outcomes, preserving process coupling and traceability;
(3) a general querying module that supports natural-language exploration and visualization over the constructed graph for cross-paper analysis;
(4) an evaluation on $\sim$800 peer-reviewed papers show the effectiveness 
of AgentCAT.

\section{Limitations and Ethical Considerations}
\noindent{\bf Ethical considerations.}
This work analyzes publicly available, peer-reviewed publications and does not involve human participants or sensitive data; institutional ethics review is not required.
The primary ethical consideration is intellectual property.
AgentCAT extracts structured experimental facts and stores provenance signals to support verification, but does not redistribute copyrighted text.
To respect publisher copyrights, we do not enable wholesale redistribution of source content.
Any use of extracted data should comply with publisher licenses and favor sharing structured fields rather than verbatim excerpts.
Automated extraction may contain errors; we mitigate this via evidence grounding, traceability, and review-driven correction, but outputs should be treated as decision support rather than ground truth.

\noindent{\bf Limitations.}
First, interactive querying can be time-consuming due to multi-step planning and query repair.
Second, some fine-grained misalignment and cross-section linking errors may persist in challenging cases, especially when evidence is implicit or distributed across modalities.
Third, our evaluation focuses on an energy-catalysis corpus; generalization to other chemical engineering subdomains may require further schema refinement and validation.

\section{GenAI Disclosure}
Generative AI tools were used solely to improve the clarity and conciseness of the manuscript's language (e.g., grammar, wording, and style). All scientific contributions, including the research ideas, methodology, experimental design, implementation, data collection/processing, results, and conclusions, were produced and verified by the authors. No Generative AI tool was used to generate or alter experimental results, figures, tables, or to produce new scientific claims.

\bibliographystyle{ACM-Reference-Format}
\bibliography{Main}
\appendix
\section{Additional Evaluation Results}
\begin{figure}[htb]
\centering
\includegraphics[width=0.82\linewidth]{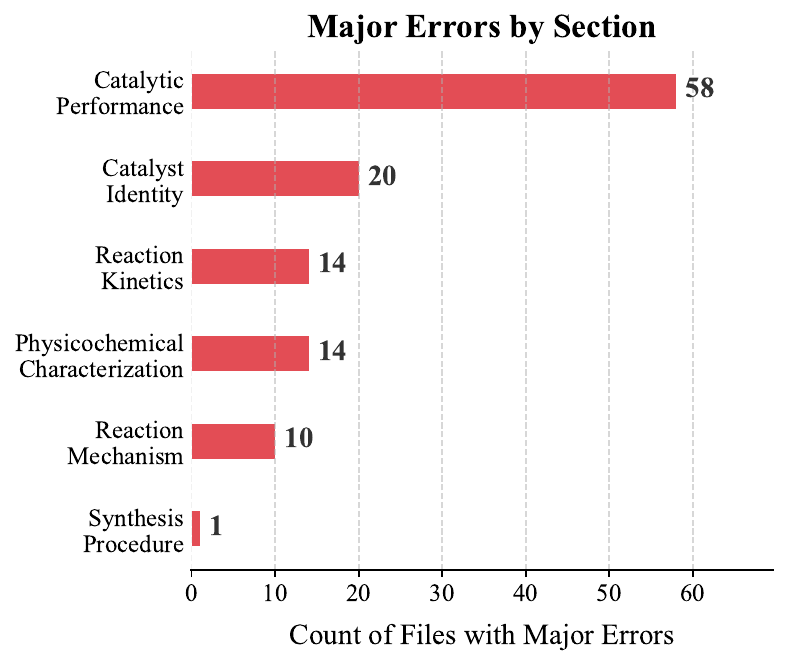}
\caption{Breakdown of \texttt{MAJOR\_ERROR} cases by section type.}
\label{fig:majorerror} 
\end{figure}

This appendix provides additional evidence for Section~\ref{sec:53}. Table~\ref{tab:Expert Evaluations} reports blind-review results for nine papers, including scores on \textit{Completeness}, \textit{Accuracy}, and \textit{Readability}, plus expert comments. Figure~\ref{fig:majorerror} breaks down \texttt{MAJOR\_ERROR} cases by section type. Most errors occur in \textit{Catalytic Performance} (58), followed by \textit{Catalyst Identity} (20), while \textit{Synthesis Procedure} has very few (1). These results contextualize the aggregate scores and guide pipeline refinement.
\label{apd:result}
\begin{table*}[!ht]
    \centering
    \caption{Expert Evaluations of Nine Representative Papers (selected from the 20 evaluated papers)} 
    \label{tab:Expert Evaluations}
    \begin{tabularx}{\textwidth}{
  >{\raggedright\arraybackslash}p{0.27\textwidth}   
  >{\raggedright\arraybackslash}p{0.08\textwidth}          
  >{\raggedright\arraybackslash}X                   
}
        \toprule
        \textbf{Paper Title} & \textbf{Scores} & \textbf{Expert Evaluation} 
        \\
        \midrule
        \textbf{[Cu$_2$O]$^{2+}$ Active Site Formation in Cu-ZSM-5 Geometric and Electronic Structure Requirements for N$_2$O Activation} &
        9.3/9.7/9.3 &
    Recommendation: For industrial catalysts, long-term stability (life test) is more important than single-point data. It is recommended to add a "long\_term\_stability" field to extract key conclusions...
        \\
        
        \midrule
        \textbf{A computational and conceptual DFT approach to the kinetics of acid zeolite catalyzed electrophilic aromatic substitution reactions} &
        8.6/9.6/9 &
         The core research object of the original paper is "electrophilic substitution of aromatics (e.g., benzylation)". Although the rate constants of aromatic reactions (data in Table 1) are not extracted into catalytic\_performance (possibly because they are calculated values rather than experimental values), they are correctly placed in the energy barrier of reaction\_mechanism.
        \\
        
        \midrule
        \textbf{A Mechanistic Study of the Brønsted-Acid Catalysis of n-Hexane to Propane + Propene, Featuring Carbonium Ions} &
        9.3/8.6/9.6 &
        JSON uniformly converts all energy data (e.g., 16.23 kcal/mol to 0.704 eV). While physically correct and convenient for cross-reference comparisons, it is recommended to retain the original units or use the 0.704 eV (16.23 kcal/mol) format when tracing data sources. Older literature in chemical engineering and catalysis often uses kcal/mol, and direct conversion may cause momentary confusion for experts during verification. 
        \\
        
        \midrule
        \textbf{1,2,4-Trimethylbenzene disproportionation over large-pore zeolites: An experimental and theoretical study} &
        9.3/9/9.6 &
        The extraction quality is extremely high, especially in the depth of structural analysis of the reaction mechanism. It is recommended to supplement the research objectives and core conclusions regarding the "shape-selectivity of the LaNa-MOR transition state," optimize the organization of the acidity data, and move the reaction pathway description from reaction\_kinetics to reaction\_mechanism.
        \\

        \midrule
        \textbf{A Periodic Density Functional Theory Analysis of Direct Methane Conversion into Ethylene and Aromatic Hydrocarbons Catalyzed by Mo$_4$C$_2$-ZSM-5} &
        9.3/9.6/10 &
        The extraction comprehensively covers the core reaction mechanism, key energy barriers, and computational models. The structure is clear and the pathway is complete. However, some theoretical details (such as free energy analysis and model selection criteria) and adsorption data are not fully included. For DFT studies, the key energy information is accurately extracted.
        \\

        \midrule
        \textbf{A New Mechanism for the First Carbon-Carbon Bond Formation in the MTG Process: A Theoretical Study} &
        9.3/9/10 &
         The JSON structure is clear and highly readable. The energy barrier units have been converted from kcal/mol in the original text to eV, and the numerical correspondences are generally accurate; however, the precise correspondence of energy values requires careful verification.
        \\

        \midrule
        \textbf{A DFT Study for Catalytic Deoxygenation of Methyl Butyrate on a Lewis Acid Site of ZSM-5 Zeolite} &
        9.3/8.6/9.3 &
        The extraction results are excellent overall, accurately capturing the core of this DFT study: the ZSM-5 Lewis acid model, seven different bond breaking pathways ($\alpha$\ / \ $\beta$, C-C, C-O), detailed energy barrier data, and optimal pathway conclusions. The data structure is clear. 
        \\

        \midrule
        \textbf{$^{13}$C MAS NMR mechanistic study of the initial stages of propane activation over H-ZSM-5 zeolite} &
        9.6/9.6/9.6 &
        The extracted data is very complete and accurate, clearly distinguishing various reaction pathways and mechanisms under different catalyst acidity, propane loading, and additives (C$_3$H$_6$, H$_2$, etc.). It is recommended to supplement the "catalytic\_performance" section with initial selectivity or quantitative reaction rate data for key products, or to add a "key\_findings" field at the top level to summarize core conclusions such as "strong proton site dominance" and "pressure-dependent path switching."
        \\

        \midrule
        \textbf{A $^{13}$CO isotopic study on the CO promotion effect in methane dehydroaromatization reaction over a Mo/HMCM-49 catalyst} &
        9.3/9.6/9.6 &
        The extracted content is comprehensive and accurately describes the core findings of the catalyst, synthesis, CO-promoting mechanism, and isotope labeling experiments. It is recommended to supplement the "catalytic\_performance" section with stability information such as catalyst deactivation rate and induction period duration, or to add a "key\_conclusions" field at the top level to extract the key conclusion that "CO dissociation and CH$_4$ decomposition are independent processes that interfere with each other through the carbon pool."
        \\
        
        \bottomrule
    \end{tabularx}
\end{table*}

\section{Supplementary Material}

The literature \textbf{dataset, expert rating sheets, and schema file} are publicly available and can be accessed via the following link:
\url{https://osf.io/xrtq2/overview?view_only=f068119f2bdb483fa953790d51b7ff5b}.
\noindent
The \textbf{code implementation} is available at:
\url{https://anonymous.4open.science/r/agentcat-E69B}.

\begin{figure*}[!ht]
  \centering
  \includegraphics[width=\textwidth]{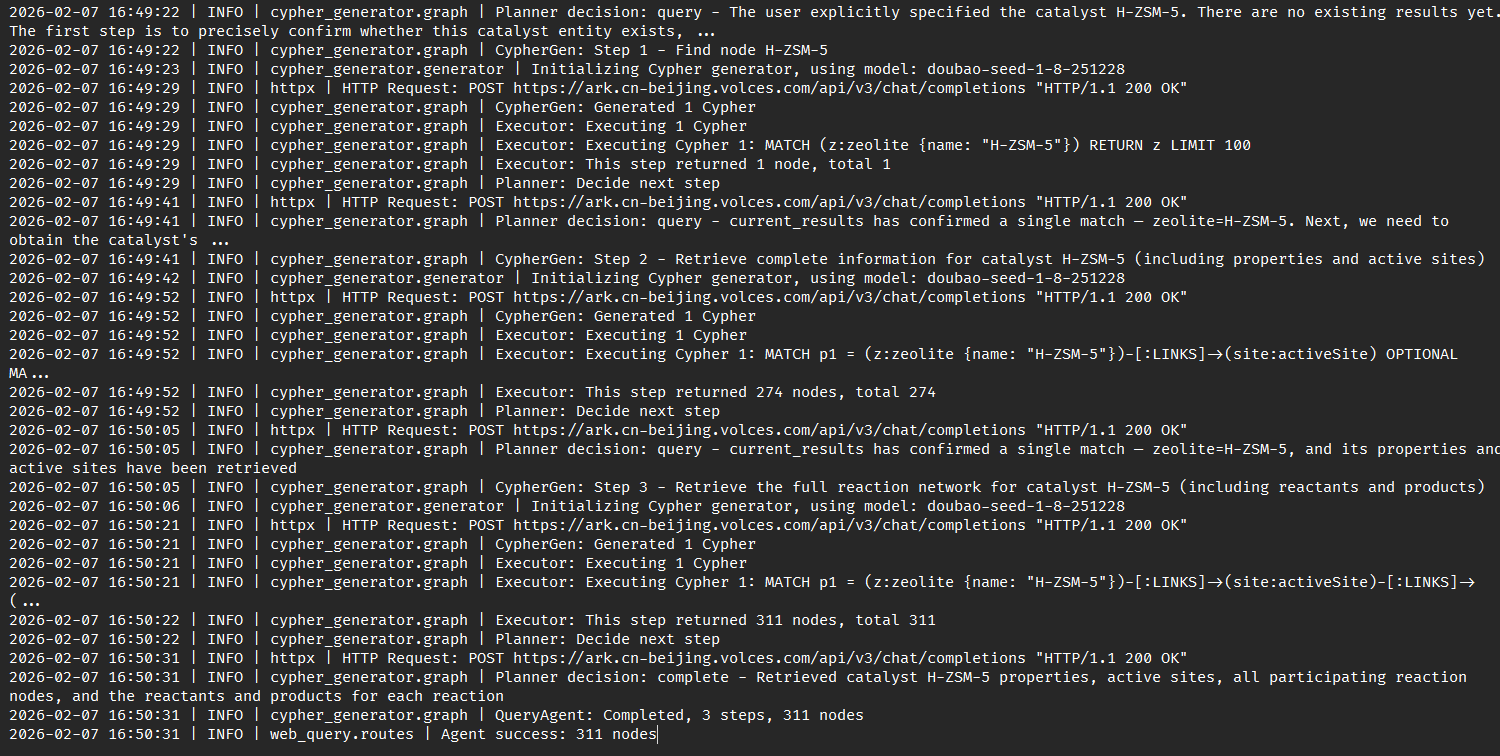} 
  \caption{Backend logs showing the General Querying Agent's query interpretation and task decomposition process.}
  \label{fig:log}
\end{figure*}

\section{User Interface}
\label{apd:sys_demostrat}
\begin{figure*}[!ht]
  \centering
  \begin{subfigure}{\textwidth}
    \centering
    \includegraphics[width=\linewidth]{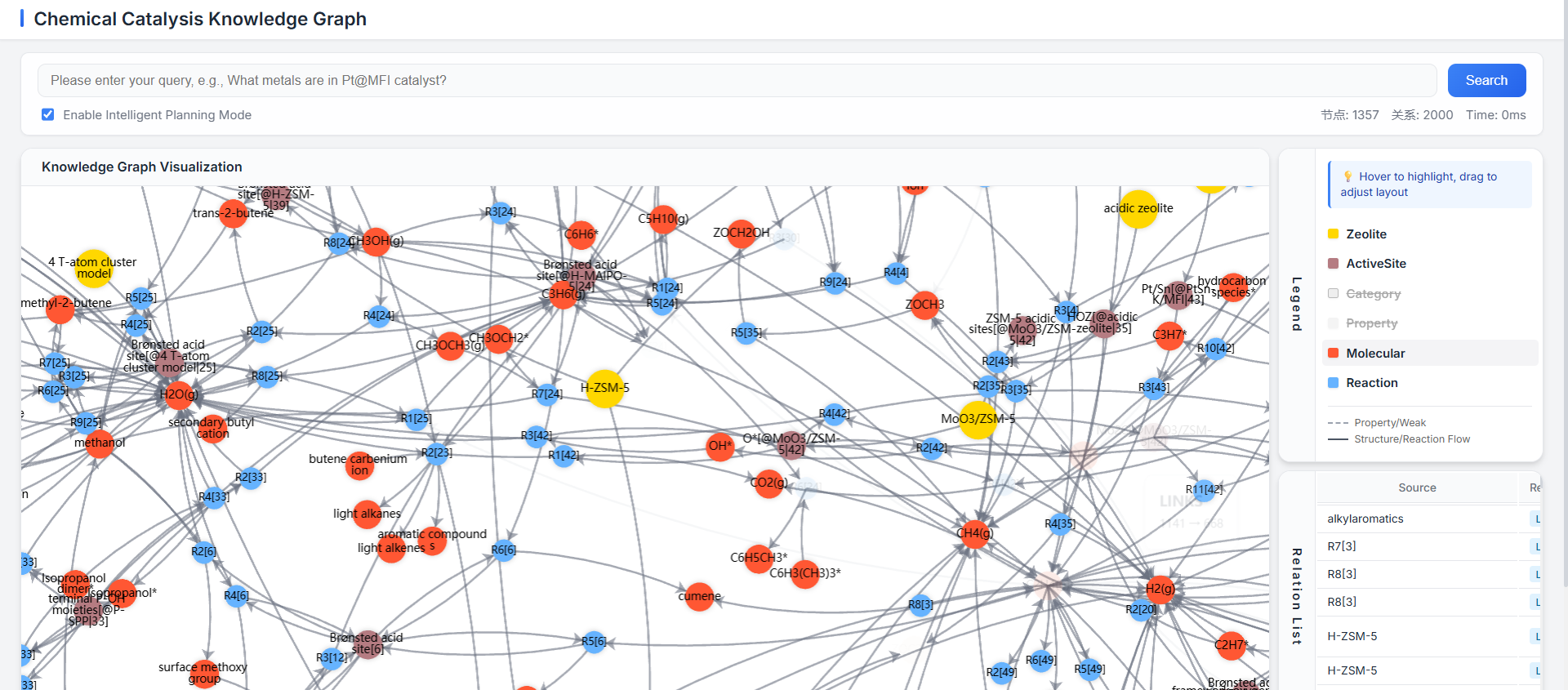}
    \caption{Default interface displaying the full reaction network.}
    \label{fig:interface_default}
  \end{subfigure}
  \vspace{0.5em}
  \begin{subfigure}{\textwidth}
    \centering
    \includegraphics[width=\linewidth]{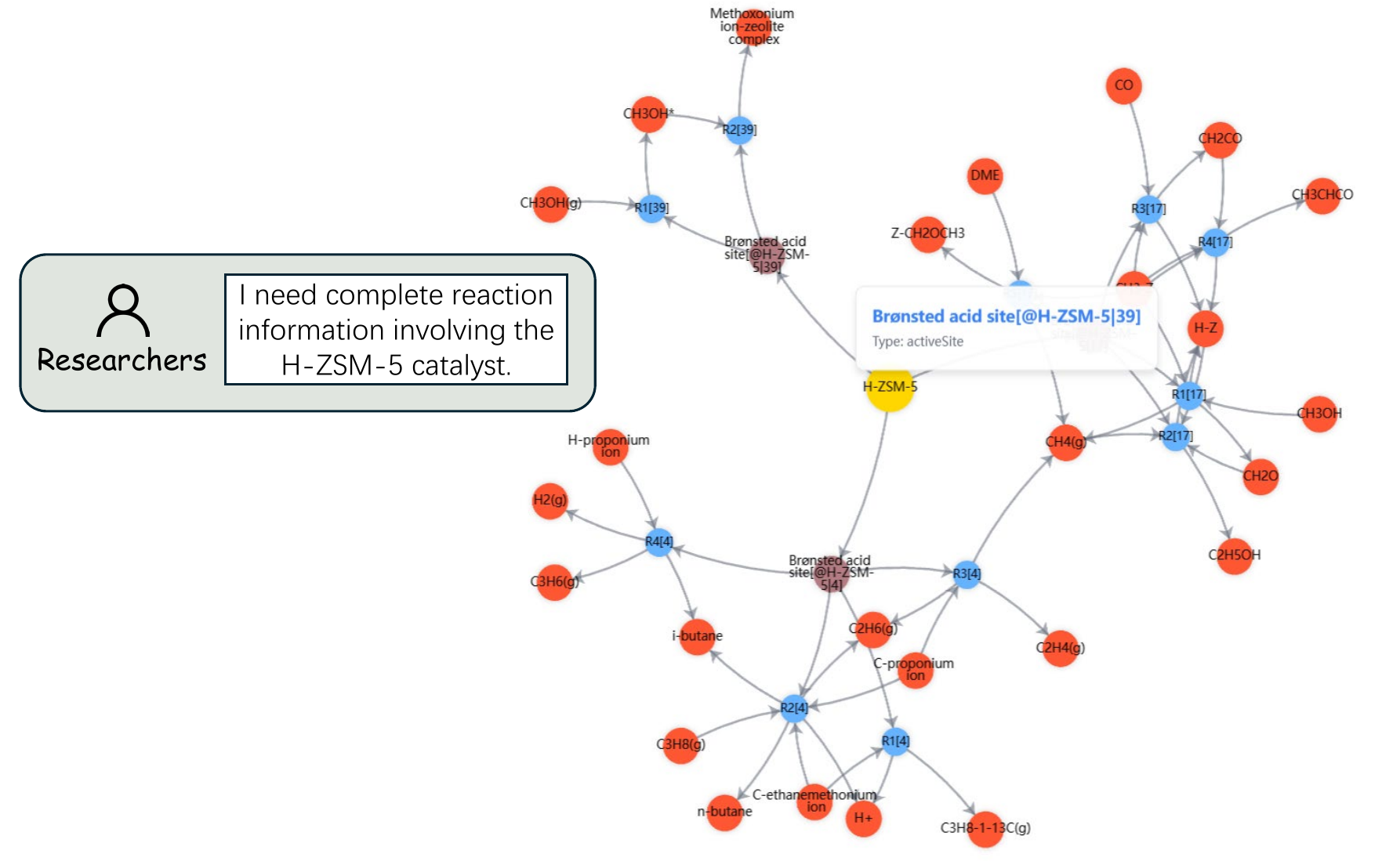}
    \caption{Query results returned by the General Querying Agent. Catalyst properties are hidden for visual clarity.}
    \label{fig:interface_query}
  \end{subfigure}
  \caption{The researcher-facing visualization interface of AgentCAT.}
  \label{fig:interface}
\end{figure*}

Figure~\ref{fig:interface} illustrates the researcher-facing interface of AgentCAT. Figure~\ref{fig:interface_default} displays the default view, showing the complete reaction network retrieved from the knowledge graph. Figure~\ref{fig:interface_query} demonstrates a natural language query submitted by the researcher, along with the results returned by the General Querying Agent. Note that in the query result visualization, catalyst property details are hidden to reduce visual clutter and improve readability. The backend logs capturing how the agent interprets and decomposes the query task are shown in Figure~\ref{fig:log}.
\end{document}